\documentclass[letterpaper,prl,twocolumn,showpacs,superscriptaddress]{revtex4}

\usepackage{graphicx}
\usepackage{subfigure}
\usepackage{bm}
 
\begin{document}
\title
{From random walk to single-file diffusion}

\author{Binhua Lin}
\email{b-lin@uchicago.edu}
\affiliation{The James Franck Institute, Department of Chemistry and CARS,
The University of Chicago, Chicago, IL 60637}

\author{Mati Meron} 
\affiliation{The James Franck Institute, Department of Chemistry and CARS,
The University of Chicago, Chicago, IL 60637}

\author{Bianxiao Cui} 
\altaffiliation[Present address: ]
{Department of Physics, Stanford University,
Stanford, California 94305}
\affiliation{The James Franck Institute, Department of Chemistry and CARS,
The University of Chicago, Chicago, IL 60637}

\author{Stuart A.\ Rice}
\affiliation{The James Franck Institute, Department of Chemistry and CARS,
The University of Chicago, Chicago, IL 60637}

\author{Haim Diamant}
\affiliation{School of Chemistry, Raymond and Beverly Sackler Faculty
of Exact Sciences, Tel Aviv University,
Tel Aviv 69978, Israel}

\date{\today}

\begin{abstract}
We report an experimental study of diffusion in a quasi-one-dimensional (q1D) colloid suspension which behaves like a Tonks gas.  The mean squared displacement as a function of time is described well with an ansatz encompassing a time regime that is both shorter and longer than the mean time between collisions.  This ansatz asserts that the inverse mean squared displacement is the sum of the inverse mean squared displacement for short time normal diffusion (random walk) and the inverse mean squared displacement for asymptotic single-file diffusion (SFD) ($\frac{1}{<x^2(t)>}=\frac{1}{2D_{o}t}+\frac{1}{2Ft^{1/2}}$ where $D_o$ is the q1D self-diffusion coefficient and $F$ is the single-file 1D mobility).  The dependence of $F$ on the concentration of the colloids agrees quantitatively with that derived for a hard rod model, which confirms for the first time the validity of the hard rod SFD theory.  We also show that a recent SFD theory by Kollmann \cite{970} leads to the hard rod SFD theory for a Tonks gas.
\end{abstract}

\maketitle

The diffusion of particles in quasi-one-dimensional (q1D) pores and channels is a basic feature of ion transport in cell membranes, molecular motion in zeolites, and particle flows in microfluidic devices (see references in \cite{879}).  The unique feature that separates q1D diffusion from diffusion in higher dimensions is the geometric confinement that forces the particles into a single file with a fixed spatial sequence.  This confinement generates a self-diffusion mechanism that has different time dependences of the mean squared particle displacement in different time domains.    

For time intervals shorter than the time between particle collisions, in the presence of a randomizing background fluid (e.g. a colloid particle in a solvent), the probability density for the particle displacement is
\begin{equation}
P_{S}(x,t)=\frac{1}{\sqrt{4\pi D_{o}t}}\exp{\Bigg\{-\frac{x(t)^{2}}{4D_{o}t}}\Bigg\}, 
\label{eq:short t prob}
 \end{equation}
where $x$ is the displacement during time interval $t=t_{1}-t_{0}$, and $D_{o}$ is the q1D self-diffusion coefficient.  However, the fixed spatial sequence of the particles severely restricts the possibility for large single particle displacements and, therefore, drastically reduces the diffusion rate at long time.  An analytic description of 1D diffusion in a system of hard rods with stochastic background forces was first reported by Harris  \cite{978}.  Several other 1D systems have been examined with a similar approach \cite{880,882,881,883,856,857,854}; the results obtained converge to the same solution.  For an infinite 1D system the long time behavior of the probability density for displacement is 
\begin{equation}
P_{L}(x,t)=\frac{1}{\sqrt{4\pi Ft^{1/2}}}\exp{\Bigg\{-\frac{x(t)^{2}}{4Ft^{1/2}}}\Bigg\},
\label{eq:long t prob}
 \end{equation}
where $F$ is a 1D mobility defined by
\begin{equation}
F=F^{HR}=l\sqrt{\frac{D_{o}}{\pi}}=\frac{1-\rho\sigma}{\rho}\sqrt{\frac{D_{o}}{\pi}}=D_{o}\sqrt{\frac{2t_{c}}{\pi}}.
\label{eq:SFD}
\end{equation}
We denote the 1D mobility of the hard rods by $F^{HR}$.  In Eq.(\ref{eq:SFD}) $\sigma$  is the particle length, $\rho$ is the 1D number density, $l$ is the mean spacing between the particles, and $t_{c}=l^2/2D_{o}$ is the mean time between collisions in the system.  Equations (\ref{eq:long t prob}) and (\ref{eq:SFD})  draw a remarkably simple picture of 1D diffusion at long time: the self-diffusion process, determined by the width of the probability density, is proportional to $t^{1/2}$ (i.e. $<x(t)^2>\sim t^{1/2}$) , and the proportionality constant is determined by the short time \textit{individual particle} dynamics.  

Recently Kollmann reported an analysis of the long time behavior of 1D diffusion that is valid both for atomic and colloid systems \cite{970}.  For colloid systems he finds the asymptotic particle density function displayed in Eq.(\ref{eq:long t prob}) with the 1D mobility, denoted by $F^q$, 
\begin{equation}
F^q=\frac{S(q)}{\rho}\sqrt{\frac{D_{c}(q)}{\pi}}\Bigg|_{q  \ll 4\pi/\sigma},
\label{eq:Fq}
\end{equation}
where $q$, $S(q)$, $D_{c}(q)$  are the momentum transfer, static structure factor, and the short time \textit{collective}-diffusion coefficient in $q$-space, respectively. The small $t$, small $q$ approximation for the dynamic structure factor, $S(q,t)$ \cite{907} , yields the relation
\begin{equation}
S(q,t)=S(q)\exp{\Bigg\{-q^{2}D_{c}(q)t}\Bigg\}\Bigg|_{t \ll t_{c},q  \ll 4\pi/\sigma}.
\label{eq:sqt}
\end{equation}
Kollmann's analysis predicts that the long time character of 1D diffusion is determined by the short time \textit{collective} dynamics of the system. 

Although theoretical analyses of 1D diffusion have been reported for the past four decades, the first experimental studies were reported only in the past decade.  Studies of molecular diffusion in zeolites, and of colloid particles confined in a channel lead to the result $<x(t)^2>\sim t^{1/2}$ at long time \cite{879,868,946}.  Very recently, Lutz, Kollmann and Bechinger \cite{979} reported the results of an experimental study of single-file diffusion in a strongly interacting colloid suspension.  The 1D mobility, determined from $<x(t)^2>=2Ft^{1/2}$ at long time, agrees with that determined from Eq.(\ref{eq:Fq}) at short time, as predicted \cite{970}.  However, the 1D mobility they find is only qualitatively similar to $F^{HR}$.    

The main difficulty encountered in the study of single-file diffusion is to obtain data at long time; this difficulty is most pronounced for low concentration samples.  To obtain the required data one needs a long-lived experimental system and stable instruments, such as those cleverly devised for the studies reported in references \cite{879} and \cite{979}.  

In this Letter, we report an experimental study of q1D diffusion in a weakly interacting colloid suspension confined in a narrow straight groove.  We establish an ansatz that accurately approximates the q1D diffusion process from the short time region to the long time region, thereby allowing us to study the long time single-file diffusion within a reasonable time frame (requiring a sample lifetime of $\sim$1 hour), as well as diffusion in the cross-over time region.  The experimentally determined q1D mobility of the system agrees quantitatively with $F^{HR}$.  

We note that Kollmann states that $F^q$ is not equivalent to $F^{HR}$, and the experimental results in \cite{979} support this statement.  However, we show that these two theories are equivalent when applied to a system, such as ours, which obeys the Tonks equation of state \cite{982}.     

Our experimental system consists of silica colloid spheres (density $2.2 g/cm^3$) suspended in water and confined in straight and narrow grooves.  The grooves are printed on a polydimethysiloxane substrate from a master pattern fabricated lithographically on a Si wafer (Stanford Nanofabrication Facility).  The small width of the groove ($<2\sigma$) prevents the spheres from passing one other, and gravity keeps them from escaping the groove.  The spheres are very weakly attractive ($<0.4k_{B}T$); the short-range attraction is derived from surface tension effects \cite{941}.   Digital video microscopy is used to directly track the time-dependent trajectories of the spheres along the groove (the motion transverse to the groove is very limited and, therefore, is not considered here).  Details relevant to sample preparation and data analysis have been described elsewhere \cite{946,941}.  
 
We have studied q1D diffusion at various colloid concentrations, characterized by a line packing fraction $\eta=\rho\sigma=N\sigma/L$, where $L$ is the length of the groove in the field of view, and  $N$ the number of spheres within $L$.   We used two different silica colloid suspensions.  For $\eta$=0.09, 0.17, 0.20, 0.38, 0.57, and 0.70 the samples had silica spheres with diameter $\sigma_{1}=1.58\mu m\pm0.04\mu m$ in a groove  that was $3\mu m\pm0.3\mu m$ wide and deep, and $2mm$ long.  For $\eta$=0.73 and 0.986 we used silica spheres with diameter $\sigma_{2}=3.7\mu m\pm0.1\mu m$ in a groove that was $5\mu m\pm0.1\mu m$ wide, $4\mu m\pm0.5\mu m$ deep, and $10mm$ long.  Care has been taken to assure that there were no blockages in the grooves.  We used the large spheres for the higher concentration samples because the small spheres could not be contained inside the grooves when $\eta>0.7$.   
 
 \begin{figure}[!h]

\includegraphics[width=3.1in,clip]{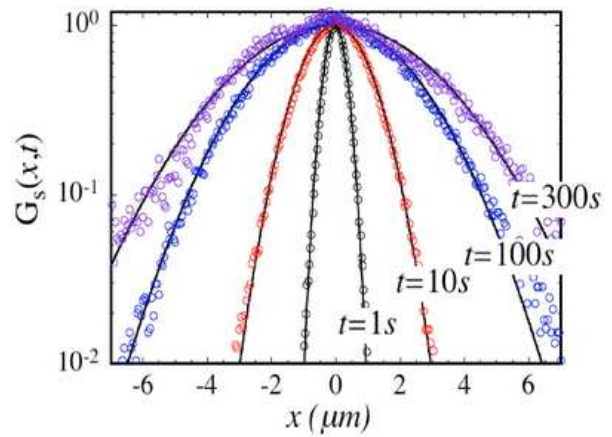} 

\caption{Typical probability density of particle displacement evolving with time (for $\eta$=0.57).  The solid lines are fits of the data to Eq.(\ref{eq:inter t prob}).}  
\label{fig:gauss}

\end{figure}

The self-diffusion process is usually described by the self-part of the van Hove function, $G_{s}(x,t)$, which is the probability density for finding a particle at a point $x_{0}+x$ at time $t_{0}+t$ given that it was at $x_{0}$ at $t_{0}$ ($G_{s}(x,t)=\frac{1}{N}\Big<\sum_{i=1}^N \delta[x+x_{i}(t_{0})-x_{i}(t_{0}+t)] \Big>$\cite{743}).  Figure \ref{fig:gauss} shows a typical $G_{s}(x,t)$ for our system, derived from time-dependent trajectories.  The deviation of $G_{s}(x,t)$ from a Gaussian, characterized by $\alpha_{2}(t)=\Big(\frac{<x(t)^4>}{3<x(t)^2>^2}-1\Big)$, is found to be negligible ($\alpha_{2}(t)\lesssim0.1$).  We therefore assume $G_{s}(x,t)$ to have the Gaussian form
\begin{equation}
G_{s}(x,t)=\frac{1}{\sqrt{2\pi <x^2(t)>}}\exp{\Bigg\{-\frac{x(t)^2}{2<x^2(t)>}}\Bigg\}.
\label{eq:inter t prob}
 \end{equation}
Figure \ref{fig:gauss} also shows the Gaussian fits to $G_{s}(x,t)$.  The mean squared displacement determined from the fitting is sensibly the same as that determined from $<x(t)^2>=\frac{1}{N}\Big<\sum_{i=1}^N [x_{i}(t_{0}+t)-x_{i}(t_{0})] ^2\Big>$.     

\begin{figure}[!h]

\includegraphics[width=3.1in,clip]{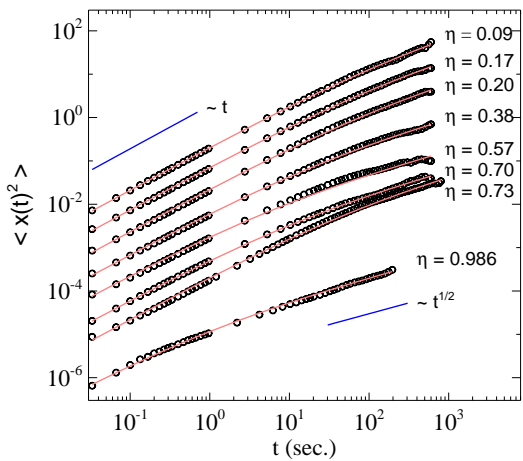}  

\caption{ Mean squared displacement as a function of $t$ at different concentrations.  Note that $<x(t)^2>$ for large spheres is scaled by the factor $\sigma_{2}/\sigma_{1}$.  The data (symbols) are shifted downward a factor of 3 from one another for clarity.  The error bars are smaller than the symbols used.  For $t\leq1s$ the movies were grabbed at $30frames/s$, and for $t>1s$ the images were grabbed at $4frames/s$ and $5frames/s$ for small and large spheres, respectively (only a subset of the data are plotted for clarity).  The solid lines are fits of the data to Eq.(\ref{eq:inter MSD}).}

\label{fig:msd}

\end{figure}

Figure \ref{fig:msd} shows  $<x(t)^2>$ as a function of $t$ at various concentrations, extracted from fitting $G_{s}(x,t)$ to Eq.(\ref{eq:inter t prob}).  Qualitatively, $<x(t)^2>$ is proportional to $t$ at short time, changes smoothly to $<x(t)^2>\sim t^{\gamma}$ ($\gamma<$1) at later time, and reaches $<x(t)^2>\sim t^{1/2}$ at long time for the higher concentrations.  Because of the expected trend in the behavior of $<x(t)^2>$ as a function of $\eta$, it is reasonable to postulate that with long enough time the low concentration samples will also exhibit $<x(t)^2>\sim t^{1/2}$.  Accordingly, we use the following ansatz to describe $<x(t)^2>$ over the entire time range: 
\begin{equation}
\frac{1}{<x^2(t)>}=\frac{1}{2D_{o}t}+\frac{1}{2Ft^{1/2}}.
\label{eq:new width}
\end{equation}
Equation (\ref{eq:new width}) leads to
\begin{equation}
<x^2(t)>=\frac{2D_{o}t}{1+(D_{0}/F)t^{1/2}} =\frac{2D_{o}t}{1+(t/t_{x})^{1/2}} .
\label{eq:inter MSD}
 \end{equation}

By construction, Eq. (\ref{eq:inter MSD}) satisfies both the short and long time limits, and it provides a characteristic cross-over time, $t_{x}=(F/D_{o})^2$.  If $F=F^{HR}$, then $t_{x}=t^{HR}_{x}\equiv2t_{c}/\pi$ (see Eq.(\ref{eq:SFD})), so that for hard rods $t_{x}$ is, essentially, the mean time between collisions.  The fits of $<x(t)^2>$ to Eq.(\ref{eq:inter MSD}) shown in Fig.\ref{fig:msd} indicate that Eq.(\ref{eq:inter MSD}) is a reasonable approximation for all time, and the fitting yields three pertinent parameters describing the q1D diffusion: the short time self-diffusion coefficient, $D_{o}$, the long time q1D mobility, $F$, and the cross-over time, $t_{x}$.  

When $\eta \leq 0.4$ the fitted values for $D_{o}$ are $D_{o1}=0.11\pm0.005 \mu m^2/s$ and $D_{o2}=0.036\pm0.005 \mu m^2/s$, respectively, for the small and large spheres (the lower concentration data for large spheres are not shown here).  These values are, within the experimental precision, the same as those calculated for isolated colloids confined by the three walls of the groove \cite{946} and, therefore, are used to determine $F^{HR}$  in Eq.(\ref{eq:SFD}) and $t_{x}$ in Eq.(\ref{eq:inter MSD}).  At higher concentrations the fitted self-diffusion coefficient is slightly smaller ($\sim 70\%-80\% D_{o})$, suggesting that hydrodynamic interaction between colloid particles comes into play even at the shortest time accessible in our experiment \cite{946, 948}.   

\begin{figure}[!h]

\includegraphics[width=3.1in,clip]{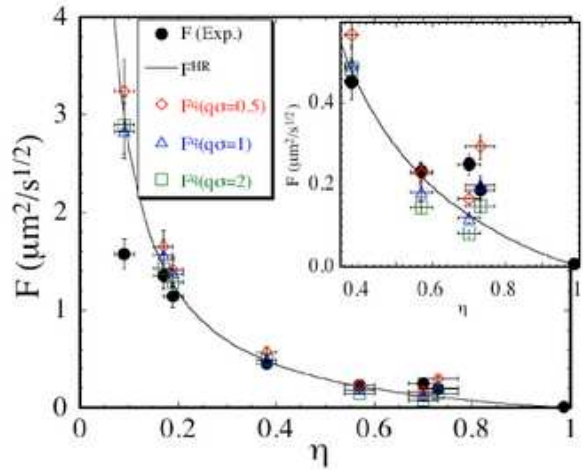}  

\caption{Quasi-1D mobility (solid circles) determined with the empirical expression (Eq.(\ref{eq:inter MSD})) as a function of concentration ($F$ for the large colloids is scaled by the factor $(\sigma_{1}/\sigma_{2})\sqrt{D_{o1}/D_{o2})}$.  The solid line represents $F^{HR}$.  Other symbols represent $F^q$ determined from Eq.(\ref{eq:Fq}).  The inset zooms into the data at higher $\eta$.}
 \label{fig:F}

\end{figure}

Figure \ref{fig:F} shows the fitted $F$ as a function of $\eta$.  When $0.17\leq\eta \leq 0.57$, $F=F^{HR}$ within the experimental precision.   However, $F\approx F^{HR}/2$ when $\eta =0.09$ and $F\lesssim 2F^{HR}$ when $\eta\geq 0.7$.  Using the fitted $F$ and $D_{o}$ we find $t_{x}=205, 150, 107, 17, 4.5, 5.2, 3.8, 0.006s$ for $\eta=0.09, 0.17, 0.19, 0.38, 0.57, 0.70, 0.73, 0.986$, correspondingly, to be compared with $t^{HR}_{x}=955, 172, 116, 19, 4.1, 1.3, 0.7, 0.001s$, respectively (note for large spheres $t_{x}$ and $t^{HR}_{x}$ are scaled by a factor $(\sigma_{1}/\sigma_{2})^{2}(D_{o2}/D_{o1})$).  For $\eta\geq0.7$ we can force $F=F^{HR}$ by replacing the colloid diameter in $F^{HR}$ with a larger effective diameter.  We speculate that at higher concentration the colloid-colloid interaction, though weak, must be accounted for.  Since the first peak of the pair correlation function is at a separation slightly larger than the sphere diameter \cite{941}, the effective sphere diameter is thereby increased.   

The accuracy of the fitted $F$ and $D_{o}$ depends on the range of $t$ relative to $t^{HR}_{x}$.  If the range of $t$ extends both to $t\ll t^{HR}_{x}$ and $t\gg t^{HR}_{x}$ we can extract $F$ and $D_{o}$ accurately from Eq.(\ref{eq:inter MSD}); if not the values obtained are less accurate, as shown by the discrepancies between the fitted $F$ and $F^{HR}$ for $\eta=0.09$, and the fitted $D_{o}$ and expected $D_{o}$ at higher $\eta$.  

\begin{figure}[!h]

\includegraphics[width=3.1in,clip]{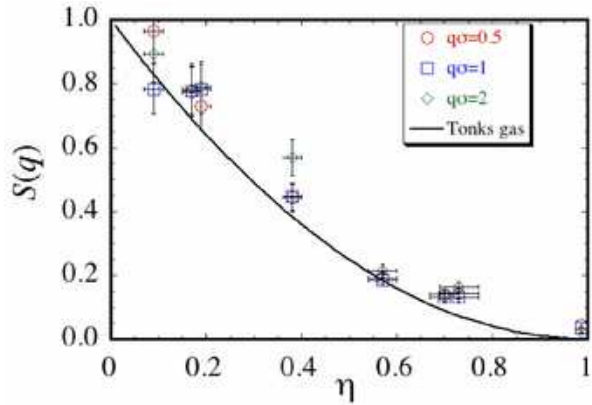}  

\caption{ The experimentally determined static structure factor $S(q)$ for small $q$ as a function of concentration, compared with that derived from the equation of state for a Tonks gas.}

\label{fig:chi}

\end{figure}

We now show that, as is to be expected, $F^{HR}=F^q$ for a q1D system that is described by Tonks equation of state $f(1-\rho\sigma)=\rho k_{B}T$ ($f$ is the linear force) \cite{982}.  The relative isothermal compressibility of a Tonks gas is
\begin{equation}
\chi_{T}/\chi_{To}=(1-\rho\sigma)^2=(1-\eta)^2=S(q)\Bigg|_{q=0}, 
 \label{eq:S0}
\end{equation}
where $\chi_{T}=-\frac{1}{L}\frac{\partial f}{\partial L}$ for 1D, and $\chi_{To}=(k_{B}T\rho)^{-1}$.  Figure \ref{fig:chi} shows $S(q)$ for our system for small $q$ ($2\pi/q \gg \sigma/2$) as a function of $\eta$; the agreement with Eq.(\ref{eq:S0}) clearly indicates that our system behaves like a Tonks gas.  The slight shift of experimental values of $S(q)$ to larger $\eta$ from that of a Tonks gas is consistent with the weak colloid-colloid attraction \cite{886}.  

Kollmann's theory relates $F^q$ to the collective diffusion coefficient, $D_{c}(q)$, and the relative isothermal compressibility, $S(0)$. Substituting $S(0)$ given in Eq.(\ref{eq:S0}) and $D_{c}(q)=D_{o}H(q)/S(q)$ \cite{907} ($H(q)$ is the hydrodynamic factor) into Eq.(\ref{eq:Fq}), we obtain $F^q=l\sqrt{\frac{D_{o}}{\pi}}=F^{HR}$, if $H(q)=1$, i.e. if hydrodynamic interaction is negligible.   

We have calculated $F^q$ as follows.  First, $S(q,t)$ was determined from the trajectories using $S(q,t) =\frac{1}{N}\Big<\rho_{q}(t)\rho_{-q}(0)\Big>$, where $\rho_{q}(t)=\frac{1}{\sqrt{N}}\int dx \exp(-iqx) \sum_{k=1}^N \delta[x-x_{k}(t)]$.  Then $S(q,t)$ was fitted to the short time approximation (Eq.(\ref{eq:sqt})) to extract $D_{c}(q)$ at small $q$ ($q\ll 4\pi/\sigma$).  Within the short time range $0.03s\leq t\leq 1s$, $S(q,t)$ is well described by Eq.(\ref{eq:sqt}) except for the case $\eta=0.986$.  Finally, $F^q$ was calculated using Eq.(\ref{eq:Fq}) for all the concentrations except $\eta=0.986$.  As shown in Fig.\ref{fig:F}, $F^q$ agrees with $F^{HR}$ within the experimental precision.  The data in Fig.\ref{fig:F} also show that $F^q$ depends on $q$, which we attribute to hydrodynamic interaction in the system.  In a q1D system hydrodynamic interaction is screened on the length scale of the channel width, so it can be treated as generating a pair-interaction \cite{948}.   Then the effect of $H(q)$ on $F^q$ is not significant.  A full discussion of $H(q)$ in the q1D system will be published separately.    

It is worth noting that  $<x(t)^2>$ in Eq.(\ref{eq:inter MSD}) is the width of a Gaussian which is the product of the short time probability density (Eq.(\ref{eq:short t prob})) and the long time probability density (Eq.(\ref{eq:long t prob})).  The success of the approximation given in Eq.(\ref{eq:inter MSD}) suggests that the van Hove function displayed in Eq.(\ref{eq:inter t prob}) is valid for all time for our system.  It is further implied that the randomizing background that determines the short time behavior and the correlated motion that determines the long time single-file diffusion are coexisting independent processes with time dependent weights.  For $t\ll t_{x}$ and $t\gg t_{x}$ the system exhibits normal diffusion and single-file diffusion, respectively.  However, for $t\sim t_{x}$, the motion of a particle in 1D is hindered by its neighbors and the short time displacement distribution is modified by the long time distribution.    

We thank Tom Witten and Sidney Nagel for helpful discussions.  This research was supported by the NSF (CTS-021174 and CHE-9977841), the Israel Science Foundation (77/03), and the NSF-funded MRSEC laboratory at The University of Chicago.



 \end{document}